\documentclass[a4paper,11pt]{article}
\pdfoutput=1
\usepackage{jcappub} 
\usepackage[latin9]{inputenc}
\usepackage{textcomp}
\usepackage{amsmath}
\usepackage{amssymb}
\usepackage{color}

\makeatletter

\newcommand{\be}[1]{\begin{equation}\label{#1}}
\newcommand{\ee}{\end{equation}}
\newcommand{\ba}[1]{\begin{eqnarray}\label{#1}}
\newcommand{\ea}{\end{eqnarray}}

\makeatother
\begin{document}

\title{Regular Instantons in the Eddington-inspired-Born-Infeld Gravity: Lorentzian Wormholes from Bubble Nucleations}
\author[a,b]{Mariam Bouhmadi-L\'{o}pez}
\author[c,d]{Che-Yu Chen,}
\author[c,d,e]{Pisin Chen,}
\author[f,g]{and Dong-han Yeom}
\affiliation[a]{Department of Theoretical Physics, University of the Basque Country UPV/EHU, P.O.~Box 644, 48080 Bilbao, Spain}
\affiliation[b]{IKERBASQUE, Basque Foundation for Science, 48011 Bilbao, Spain}
\affiliation[c]{Department of Physics and Center for Theoretical Sciences, National Taiwan University, Taipei, Taiwan 10617}
\affiliation[d]{LeCosPA, National Taiwan University, Taipei, Taiwan 10617}
\affiliation[e]{Kavli Institute for Particle Astrophysics and Cosmology, SLAC National Accelerator Laboratory, Stanford University, Stanford, CA 94305, USA}

\affiliation[f]{Asia Pacific Center for Theoretical Physics, Pohang 37673, Republic of Korea}
\affiliation[g]{Department of Physics, POSTECH, Pohang 37673, Republic of Korea}

\emailAdd{mariam.bouhmadi@ehu.eus}
\emailAdd{b97202056@gmail.com}
\emailAdd{pisinchen@phys.ntu.edu.tw}
\emailAdd{innocent.yeom@gmail.com}

\abstract{The $O(4)$-symmetric regular instanton solutions are studied within the framework of the Eddington-inspired-Born-Infeld gravity (EiBI). We find that the behavior of the solution would deviate from that in Einstein gravity when the kinetic energy of the scalar field is sufficiently large. The tunneling probability is calculated numerically in different parameter space. We find that the tunneling probability would increase with the Born-Infeld coupling constant, which is assumed to be positive in this paper. Furthermore, we discover a \textit{neck} feature in the physical instanton solutions when the kinetic energy of the scalar field is sufficiently large. This feature can be interpreted as a Lorentzian time-like wormhole geometry formed during the bubble materialization.}

\maketitle
\flushbottom

\section{Introduction}
In theoretical physics, the construction of Lorentzian wormhole solutions is always very appealing \cite{Morris:1988cz,Lorentzianwormholes}. One of the reasons is that it may replace a black hole singularity with a wormhole geometry, which is everywhere regular in spacetime. However, the wormhole structure may violate the convergence condition, that is, there exists some repulsive \textit{force} that prevents gravitational collapse. In Einstein's general relativity (GR), the violation of the convergence condition is equivalent to the violation of some energy conditions \cite{hawkingellis,Wald}. Therefore, in GR one usually needs to include the so called \textit{exotic matter}, which violates the null energy condition, to support the existence of wormhole structures. 

Instead of introducing exotic matter, one may resort to quantum corrections to GR as the justification for wormhole solutions. In fact, it is believed that GR, being an incomplete theory due to the existence of spacetime singularities \cite{Penrose:1964wq}, should be corrected by quantum effects when the curvature becomes large near the classical singularity. Until now there is still no self-consistent quantum theory of gravity. As a supplement, one may consider theories of extended gravity \cite{Capozziello:2011et}. In principle, these theories are expected to ameliorate spacetime singularities and can be viewed as effective theories of a fundamental quantum theory of gravity below some cutoff energy scale. It has been shown that in certain theories of extended gravity, wormhole solutions can exist without violating any energy condition. This can be achieved since the field equations in such theories are significantly different from those in GR and the violation of the convergence condition does not necessarily imply the violation of the energy condition \cite{Capozziello:2013vna,Capozziello:2014bqa,Lobo:2009ip,Dehghani:2009xu,Kanti:2011jz,MontelongoGarcia:2010xd,Bohmer:2011si,Bronnikov:2015pha,Bambi:2015zch}.

In this paper, we consider the Eddington-inspired-Born-Infeld gravity (EiBI) \cite{Banados:2010ix} and find the existence of a Lorentzian wormhole via quantum gravitational effects in the theory. In particular, we investigate the quantum tunneling process and the emergence of bubbles by studying the Euclidean version of the theory with a scalar field minimally coupled to gravity, that is, the Coleman-deLuccia (CDL) instanton \cite{Coleman:1980aw}. The EiBI theory was first proposed in Ref.~\cite{Banados:2010ix} and it was shown to be able to remove the big bang singularity of the universe \cite{Scargill:2012kg,Cho:2012vg,Avelino:2012ue}. The theory is equivalent to GR in vacuum but deviates from it when matter field is included. Recently, the EiBI theory has been further studied in the context of cosmology \cite{Bouhmadi-Lopez:2013lha,Bouhmadi-Lopez:2014jfa,Delsate:2012ky,EscamillaRivera:2012vz,Yang:2013hsa,Du:2014jka,Cho:2013pea,Li:2017ttl}, black holes \cite{Olmo:2013gqa,Wei:2014dka,Sotani:2014lua,Sotani:2015ewa,Chen:2018mkf}, and neutron stars \cite{Harko:2013wka,Sham:2013cya}. The constraint on the Born-Infeld coupling constant \cite{Casanellas:2011kf,Avelino:2012ge,Avelino:2012qe,Jana:2017ost}, some quantum effects \cite{Bouhmadi-Lopez:2016dcf,Albarran:2017swy}, some self-gravitating systems \cite{Roshan:2018pts}, and the singular instanton solutions \cite{Arroja:2016ffm} in the EiBI theory have also been investigated. Moreover, several interesting generalizations of the EiBI theory have been put forward in Refs.~\cite{Liu:2012rc,Makarenko:2014lxa,Fernandes:2014bka,Odintsov:2014yaa,Jimenez:2014fla,Chen:2015eha,Bouhmadi-Lopez:2017lbx,Chen:2017ify,Ping-Li:2018mrt}. See Ref.~\cite{BeltranJimenez:2017doy} for a recent review on the EiBI gravity.

Actually, the existence of Lorentzian wormholes in EiBI theory has been proposed in the literature \cite{Harko:2013aya,Shaikh:2015oha,Tamang:2015tmd,Olmo:2015bya,Olmo:2015dba,Shaikh:2018yku}. It has been proven that some wormhole solutions do not require the violation of any energy condition when the Born-Infeld coupling constant is negative, as long as the energy-momentum tensor satisfies a particular form \cite{Shaikh:2015oha,Shaikh:2018yku}. This is reminiscent of the fact that the theory admits a bouncing cosmology to replace the big bang singularity in such parameter space. However, we will show in this paper that in the EiBI theory with a positive Born-Infeld coupling constant $\kappa$, a Euclidean \textit{neck} feature can be found in the CDL instanton (or the regular instanton for distinguishing it from the singular instanton studied in Ref.~\cite{Arroja:2016ffm}) and it actually corresponds to a Lorentzian time-like wormhole after analytically continuing to the Lorentzian spacetime{\footnote{Although $\kappa$ can in principle be either positive or negative, we will only consider a positive $\kappa$ in this work because of the instability problems usually present when $\kappa$ is negative \cite{Avelino:2012ge}.}}. Similar creations of Lorentzian wormholes via quantum tunneling effects have also been studied in the context of the Gauss-Bonnet-dilaton gravity \cite{Tumurtushaa:2018agq} and scalar-tensor theories in which the scalar field is non-minimally coupled to gravity \cite{Battarra:2014naa}. 

This paper is outlined as follows. In section~\ref{sectII}, we present the equations of motion describing the regular instanton solutions in the EiBI gravity. The boundary conditions are also introduced. In section~\ref{seciii}, we first confirm the existence of the regular instanton solutions in the EiBI gravity. Then we find that in some parameter space, the solutions could possess an interesting \textit{neck} feature. The tunneling probability and its dependence on various parameters are studied in section~\ref{secprob}. In section~\ref{secvwh}, we show that the \textit{neck} feature found in section~\ref{seciii} can be interpreted as a Lorentzian wormhole which is formed during the bubble materialization. The relationship between the wormhole geometry and the energy condition is also discussed. We finally present our conclusions in section~\ref{conclu}.

\section{Instantons in the EiBI gravity}\label{sectII}
We will focus on the instanton solutions in the context of the EiBI gravity whose action is
\cite{Banados:2010ix}
\begin{equation}
S_{\mathrm{EiBI}}=\frac{1}{\kappa}\int d^4x\Big[\sqrt{|g_{\mu\nu}+\kappa R_{(\mu\nu)}(\Gamma)|}-\lambda\sqrt{-g}\Big]+S_\mathrm{M}(g).
\label{actioneibi}
\end{equation}
The theory is formulated within the Palatini variational principle in which the metric $g_{\mu\nu}$ and the affine connection $\Gamma$ are treated as independent variables. Only the symmetric part of the Ricci tensor $R_{\mu\nu}(\Gamma)$ constructed from $\Gamma$ appears in the action. Therefore, the theory has a projective symmetry and the torsion field can be simply gauged away. The expression within the square root in the first term is the absolute value of the determinant of the tensor $g_{\mu\nu}+\kappa R_{(\mu\nu)}(\Gamma)$. Furthermore, $g$ denotes the determinant of $g_{\mu\nu}$ and $S_\mathrm{M}$ stands for the matter Lagrangian, where matter is assumed to be coupled to $g_{\mu\nu}$ only. In addition, $\lambda$ is a dimensionless constant which relates to an effective cosmological constant of the theory at the low curvature limit via $\Lambda\equiv(\lambda-1)/\kappa$. The parameter $\kappa$ is the Born-Infeld coupling constant characterizing the theory. In the limit of $\kappa\rightarrow 0$, the theory reduces to GR. Note that we will work with the reduced Planck units $8\pi G=1$ and set the speed of light to unity.

In Ref.~\cite{Delsate:2012ky}, it has been shown that the EiBI action can be transformed to its Einstein frame in which the alternative action reads
\begin{equation}
S_{\mathrm{EiBI2}}=\frac{1}{2}\int d^4x\sqrt{-q}\Big[R[q]-\frac{2}{\kappa}+\frac{1}{\kappa}\Big(q^{\alpha\beta}g_{\alpha\beta}-2\sqrt{\frac{g}{q}}\lambda\Big)\Big]+S_\mathrm{M}(g),
\label{actionalternative}
\end{equation}
where $R[q]$ is the curvature scalar constructed purely from an auxiliary metric $q_{\mu\nu}$. In the above action, $q^{\mu\nu}$ and $q$ represent the inverse and the determinant of $q_{\mu\nu}$, respectively. The actions \eqref{actioneibi} and \eqref{actionalternative} are dynamically equivalent in the sense that they give the same equations of motion. Note that the EiBI field equations can be obtained by varying the action \eqref{actionalternative} with respect to the metrics $g_{\mu\nu}$ and $q_{\mu\nu}$ separately. According to the equations of motion, the auxiliary metric $q_{\mu\nu}$ can be written as $q_{\mu\nu}=g_{\mu\nu}+\kappa R_{(\mu\nu)}$ and it turns out to be compatible with the affine connection $\Gamma$. 

In Refs.~\cite{Bouhmadi-Lopez:2016dcf,Arroja:2016ffm,Albarran:2017swy} where the quantum effects in the EiBI theory are investigated, the action \eqref{actionalternative} has been shown to be more convenient, for example, to construct the Hamiltonian of the system \cite{Bouhmadi-Lopez:2016dcf,Albarran:2017swy}, or to deduce the on-shell action when studying the instanton solutions \cite{Arroja:2016ffm}. In this work, we will as well use the action \eqref{actionalternative} to study the instanton solutions in the EiBI framework, as done in Ref.~\cite{Arroja:2016ffm}. 

\subsection{Equations of motion}
To investigate the instanton solutions within the EiBI theory, we consider the EiBI gravity minimally coupled to a scalar field $\phi$ with a potential $V(\phi)$. The Euclidean version of the action is constructed from the action \eqref{actionalternative} after Wick rotation ($x^0\rightarrow ix^0_\mathrm{E}$ and $S_\mathrm{E}=iS_{\mathrm{EiBI2}}$) \cite{Arroja:2016ffm}
\begin{align}
S_\mathrm{E}=&-\frac{1}{2}\int d^4x\sqrt{+q}\Big[R[q]-\frac{2}{\kappa}+\frac{1}{\kappa}\Big(q^{\alpha\beta}g_{\alpha\beta}-2\sqrt{\frac{g}{q}}\lambda\Big)\Big]\nonumber\\
&+\int d^4x\sqrt{+g}\Big(\frac{(\nabla\phi)^2}{2}+V(\phi)\Big)+S_{\mathrm{B}}.
\label{euaction}
\end{align}
The last term $S_\mathrm{B}$ is the boundary action which can be expressed as the Gibbons-Hawking boundary action of the auxiliary metric $q_{\mu\nu}$ \cite{Gibbons:1976ue}. Since what we are going to investigate in this work are the regular instanton solutions whose contributions to the boundary terms are zero due to the nonsingular boundary conditions, we will omit the boundary term $S_\mathrm{B}$ in the rest of this paper.

We consider $O(4)$-symmetric instantons which in general have dominating contributions in the path integral formulation of quantum gravity. These instantons can be described by the following Euclidean minisuperspace metrics:
\begin{align}
ds_{g}^{2} &= N^{2}(\tau) d\tau^{2} + a^{2}(\tau) d\Omega_{3}^{2}\,,\\
ds_{q}^{2} &= M^{2}(\tau) d\tau^{2} + b^{2}(\tau) d\Omega_{3}^{2}\,,
\end{align}
where $d\Omega_3^2$ is the metric on a three-sphere. In the above equations, $a$ and $b$ ($N$ and $M$) are the scale factors (lapse functions) of the physical and auxiliary metrics, respectively. These functions are all functions of the Euclidean time $\tau$ in our setup.

After varying the action, one can obtain the constraint equations as follows \cite{Arroja:2016ffm}:
\begin{align}
\frac{b^{4}}{a^{4}} &= \left( \lambda + \kappa V(\phi) \right)^{2} - \frac{\kappa^{2} \dot{\phi}^{4}}{4 N^{4}},\label{con3}\\
\frac{M^{4}}{N^{4}} &= \frac{\left( \lambda + \kappa V(\phi) + \frac{\kappa \dot{\phi}^{2}}{2 N^{2}} \right)^{4}}{\left( \lambda + \kappa V(\phi) \right)^{2} - \frac{\kappa^{2} \dot{\phi}^{4}}{4 N^{4}}},\label{con4}\\
\frac{\dot{b}^2}{M^2}&=1-\frac{b^2}{3\kappa}\Big(1+\frac{N^2}{2M^2}-\frac{3a^2}{2b^2}\Big),\label{confirst}
\end{align}
where the dot denotes the derivative with respect to $\tau$. Furthermore, the equations of motion describing the dynamics of the system are:
\begin{align}
\ddot{\phi} &= - \left( 3 \frac{\dot{a}}{a} - \frac{\dot{N}}{N} \right) \dot{\phi} + N^{2} V', \label{eq:ddphi}\\
\ddot{b} &= \frac{\dot{M}}{M} \dot{b} - \frac{M^{2}}{3\kappa} b + \frac{N^{2}}{3\kappa} b,\label{eqddb}
\end{align}
where the prime denotes the derivative with respect to the argument, i.e., $V'=dV/d\phi$. The instanton solutions can be derived by solving the differential equations \eqref{eq:ddphi} and \eqref{eqddb}, in which all variables other than $b$, $\phi$, and their derivatives should be eliminated by using Eqs.~\eqref{con3} and \eqref{con4}. Note that the constraint equation \eqref{confirst} is a redundant equation and one needs to check the fulfillment of this constraint after deriving the solutions.

Finally, the on-shell action can be obtained by inserting the above field equations into the Euclidean action \eqref{euaction}. The result is \cite{Arroja:2016ffm}
\begin{align}
S_{\mathrm{E}} &= \frac{2\pi^{2}}{\kappa} \int d\tau M b^{3} \left( \frac{a^{2}}{b^{2}} - 1\right)\label{2.11}\\
&= \frac{2\pi^{2}}{\kappa} \int d\tau Na^{3} \left( \lambda + \kappa V(\phi) + \kappa \frac{\dot{\phi}^{2}}{2N^2} \right) \left( 1 - \frac{b^{2}}{a^{2}} \right).\label{2.12}
\end{align}

\subsection{Reducing auxiliary fields}

From now on, we choose $N = 1$ without loss of generality. In order to solve $\phi$ and $b$, we need to replace $M$, $\dot{M}$, $a$, and $\dot{a}$ with $\phi$, $b$ and their derivatives by using the constraint equations \eqref{con3} and \eqref{con4}. 

First, one can get $a$ and $M$ with equations \eqref{con3} and \eqref{con4}:
\begin{align}
a &= \frac{b}{\left[\left( \lambda + \kappa V(\phi) \right)^{2} - \kappa^{2} \dot{\phi}^{4}/4 \right]^{1/4}},\label{asimp}\\
M &= \frac{\lambda + \kappa V(\phi) + \kappa \dot{\phi}^{2}/2}{\left[\left( \lambda + \kappa V(\phi) \right)^{2} - \kappa^{2} \dot{\phi}^{4}/4 \right]^{1/4}}.
\end{align}
Second, we calculate $\dot{a}$ by directly differentiating \eqref{asimp}. The result turns out to be a function of $b$, $\dot{b}$, $\phi$, $\dot{\phi}$, and $\ddot{\phi}$. One can further remove $\ddot{\phi}$ term by using the right hand side of Eq.~(\ref{eq:ddphi}). After substituting $a$, we get the result:
\begin{equation}
\dot{a} = \frac{ \sqrt{2} \left( 4 \left( \lambda + \kappa V \right)^{2} - \kappa^{2} \dot{\phi}^{4} \right)^{-1/4} \left[ - \kappa b V' \dot{\phi} + \dot{b} \left( \kappa \dot{\phi}^{2} + 2 \left(\lambda +\kappa V \right) \right) \right]}{\left( \kappa \dot{\phi}^{2} + 2 \left( \lambda + \kappa V \right) \right) \left[ 1 + 3 \kappa^{2} \dot{\phi}^{4} \left( 4(\lambda + \kappa V)^{2} - \kappa^{2} \dot{\phi}^{4} \right)^{-1} \right]}.
\end{equation}
Finally, the function $\dot{M}$ can be rewritten as a function of $\phi$, $\dot{\phi}$, $b$, and $\dot{b}$ as well by using a similar procedure. After the above reduction of variables, one can derive two second order ordinary differential equations with two functions $b$ and $\phi$. Once appropriate boundary conditions are taken into account, the functions $b$ and $\phi$ for the regular instantons can be solved.

\subsection{Boundary conditions for regular instantons}
For the regular instanton solutions, the most straightforward way to solve the equations is to assume a specific potential $V(\phi)$ at the very beginning and to find the $O(4)$-symmetric solution satisfying the following boundary conditions
\begin{align}
b(0)&=0\,,\quad\phi(0)=\phi_{0}\,,\quad\dot{b}(0)=\sqrt{\lambda + \kappa V(\phi_{0})}\,,\quad\dot{\phi}(0)=0\,,\\
b(\tau_\mathrm{f})&=0\,,\quad\dot{b}(\tau_\mathrm{f})=-\sqrt{\lambda + \kappa V(\phi(\tau_\mathrm{f}))}\,,\quad\dot{\phi}(\tau_\mathrm{f})=0\,,
\end{align} 
where $\tau_\mathrm{f}$ is the other end point of the solution on which the regularity conditions $b=\dot\phi=0$ are satisfied. Note that the boundary conditions of $\dot{b}$ are required by $b(0)=b(\tau_\mathrm{f})=0$ in combination with the constraint equation \eqref{confirst}.

In practice, the technical difficulty in solving this problem is that the value of the end point $\tau_\mathrm{f}$ can only be determined after solving the equations of motion. It is impossible to insert the boundary conditions on the end point in the first place because the value of $\tau_\mathrm{f}$ is unknown. This leads to the fine-tuning issue of the value of $\phi_0$ in the sense that the solutions may not satisfy the boundary conditions at the end point if $\phi_0$ is chosen arbitrarily. One needs to insert a fine-tuned $\phi_0$ such that the solution fulfills the boundary conditions at the end point. 

In the first subsection of the next section, we will firstly use this method to derive the regular instanton solutions in the EiBI gravity model. Explicitly speaking, we will assume a specific double-well potential in the beginning and solve the differential equations after inserting the aforementioned boundary conditions (including the fine-tuned value of $\phi_0$). We will show that the regular instanton solutions obeying the boundary conditions do exist as long as an appropriate $\phi_0$ is assumed. After proving the existence of regular instantons in the EiBI gravity, we will use a different method to derive the solutions and demonstrate how the fine-tuning issue can be skirted. 

\section{Solution analysis}\label{seciii}
In this section, we will introduce two different methods to derive the regular instanton solutions in the EiBI gravity. The first method, which will be presented in subsection~\ref{firstmethod}, consists in assuming a particular double-well potential as a prior. Then we choose $\phi_0$ such that the solution satisfies the boundary conditions mentioned in the previous section. The result justifies the existence of regular instanton solutions in the EiBI gravity. However, due to the fine-tuning issue of $\phi_0$, this method is rather difficult to be applied further given that what we want is to check how the solutions of the model depend on the parameter space. Therefore, in subsection~\ref{secondmethod}, we will use an alternative method to avoid the numerical complexity resulting from the fine-tuning issue.
\subsection{Coleman-deLuccia instantons in a double-well potential}\label{firstmethod}
\begin{figure}
\begin{center}
\includegraphics[scale=0.7]{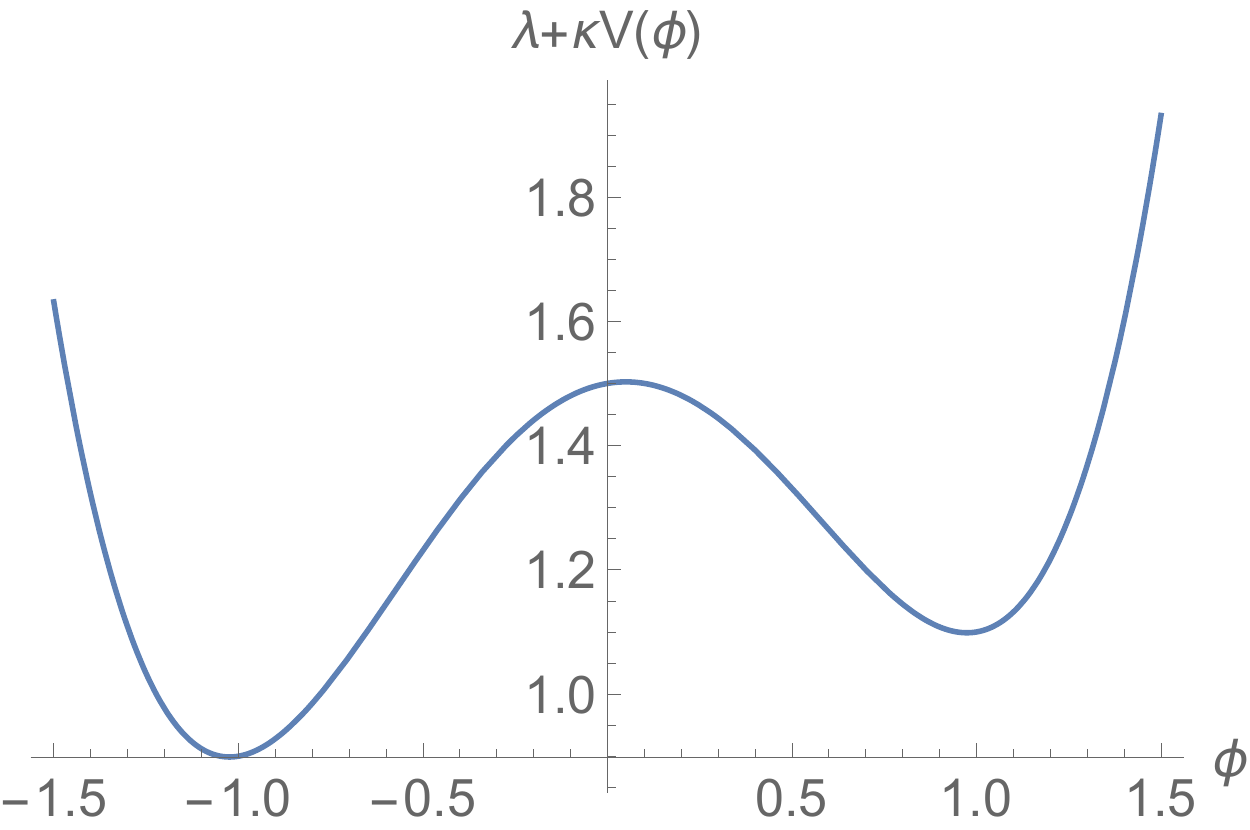}
\includegraphics[scale=0.7]{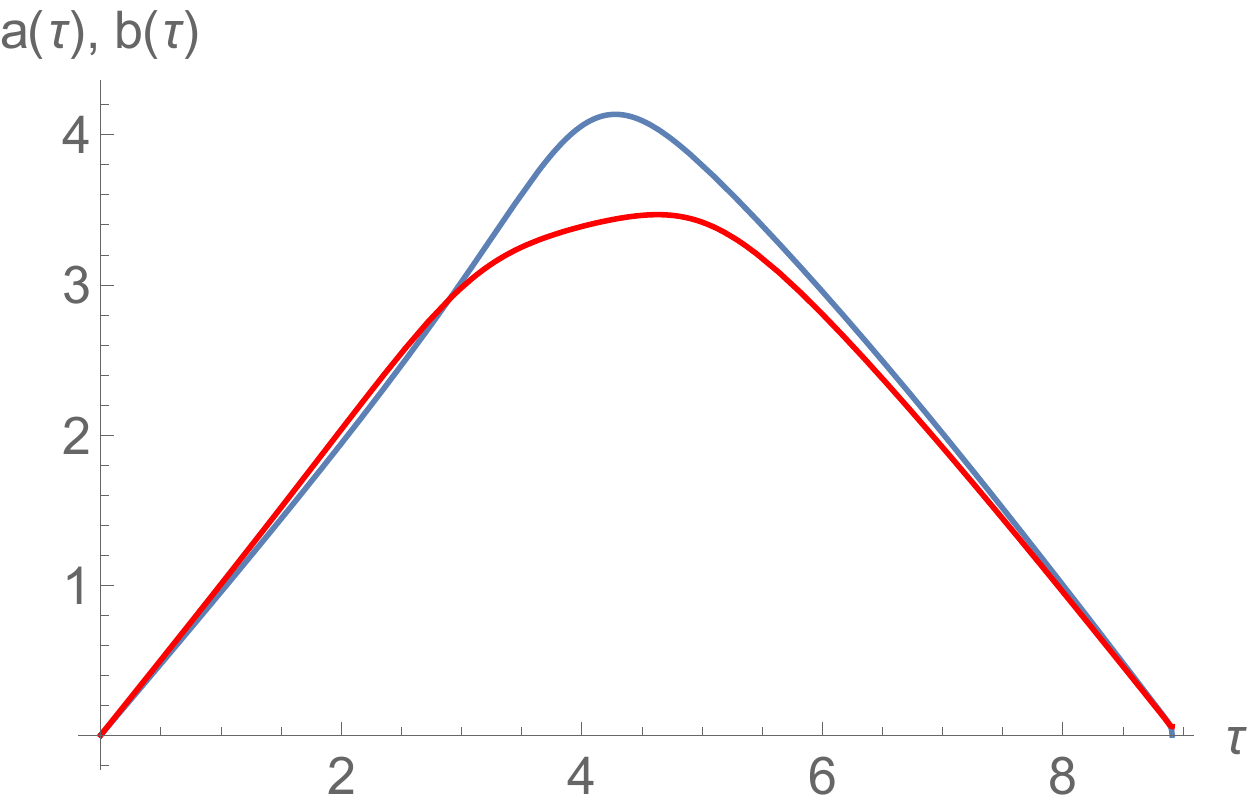}
\includegraphics[scale=0.7]{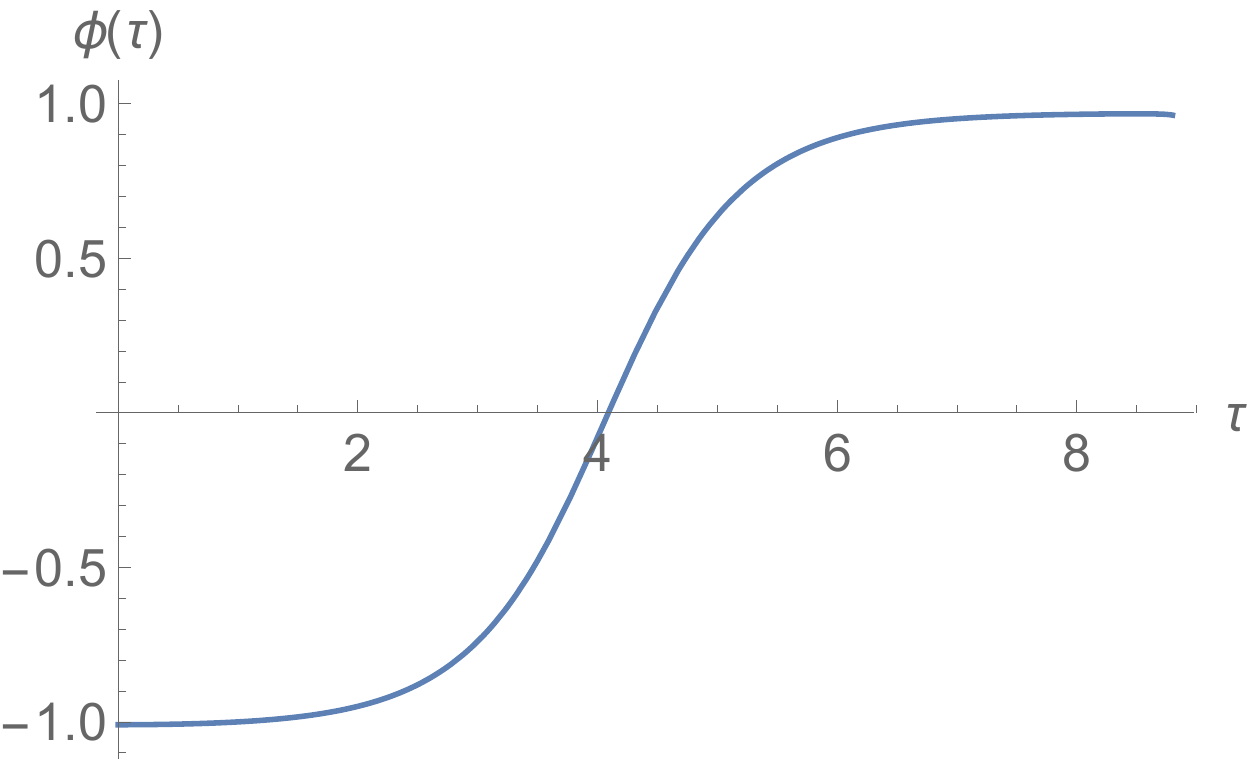}
\caption{\label{fig:pot1}The regular instanton solutions obtained by using the first method in which a double-well potential $V(\phi)$ is assumed. Upper: the biased double-well potential $\lambda + \kappa V(\phi)$ given in Eq.~(\ref{eq:v1}). Middle: the scale factors $a$ (red) and $b$ (blue). Lower: the scalar field $\phi$, for $\phi_{\mathrm{m}} = 1$, $\epsilon = 0.1$, $V_{0}=0.5$, and $\lambda=\kappa=1$.}
\end{center}
\end{figure}

In this subsection, we will assume a specific potential of the scalar field as follows:
\begin{equation}\label{eq:v1}
V(\phi) = V_{0} \left( \phi^{2} - \phi_{\mathrm{m}}^{2} \right)^{2} + \epsilon \phi,
\end{equation}
where $V_0$, $\phi_{\mathrm{m}}$, and $\epsilon$ are constants. This potential is essentially a biased double-well potential, which is plotted in Figure~\ref{fig:pot1}. Due to the parameter $\epsilon$, the potential has two local minima where the lower one corresponds to the true vacuum and the higher one is the false vacuum. After taking the fine-tuned $\phi_0$ into account and solving the equations of motion, we have found the existence of a regular instanton solution obeying the boundary conditions. The scale factors $a(\tau)$ (red) and $b(\tau)$ (blue), as well as the scalar field $\phi(\tau)$ are depicted in Figure~\ref{fig:pot1}.

It can be seen that near the initial point $\tau\rightarrow0$ and the end point where the scale factors vanish, the kinetic energy, which corresponds to $\dot\phi^2$, is very small and the behaviors of the two scale factors are almost identical. The theory hence reduces to GR. However, when the kinetic energy becomes large (see $\tau\approx4$ in Figure~\ref{fig:pot1}), the scale factors $a(\tau)$ and $b(\tau)$ would behave quite differently. This difference essentially characterizes the effect of the EiBI theory on the instanton geometry. 

\subsection{Reconstruction of the potential}\label{secondmethod}
As mentioned previously, solving the equations of motion by assuming a specific scalar field potential would suffer from the fine-tuning problem of $\phi_0$. To scrutinize further the properties of instanton solutions in the EiBI gravity model, we will introduce an alternative technique to solve the differential equations. It will be shown that this alternative method is rather convenient especially in handling our model with various parameters since one does not need to worry about the fine-tuning issue when numerically solving the equations. The idea is summarized as follows: Instead of assuming a particular scalar field potential, we assume a non-trivial form of the scalar field $\phi(\tau)$ at the beginning such that the boundary conditions at the end point have been guaranteed according to this assumption. Therefore, we can simply solve the differential equations to find $b(\tau)$ and the corresponding potential $V(\tau)=V(\phi(\tau))$ by considering the boundary conditions at the initial point $\tau=0$.

First, we give the form of the scalar field as follows:
\begin{equation}
\phi(\tau) = \frac{\tilde{\phi}(\tau) - \tilde{\phi}(0)}{\tilde{\phi}(\tau_\mathrm{f}) - \tilde{\phi}(0)},
\label{nontriphi1}
\end{equation}
where
\begin{equation}
\tilde{\phi}(\tau) = - c^{2} \left(\tau - \tau_{0}\right) + c \left(c^{2} + \tau_\mathrm{f} \tau_{0} - \tau_{0}^{2}\right) \tan^{-1}\Big(\frac{\tau-\tau_{0}}{c}\Big) + \frac{c^{2}\left(\tau_\mathrm{f}-2 \tau_{0}\right)}{2} \log \left[c^{2} + \left(\tau - \tau_{0}\right)^{2}\right],\label{nontriphi2}
\end{equation}
such that
\begin{equation}
\frac{d\tilde{\phi}}{d\tau} = -\frac{\tau \left( \tau - \tau_\mathrm{f} \right)}{1 + (\tau-\tau_{0})^{2}/c^{2}}.
\end{equation}
In the above definition, $\tau_0$, $\tau_\mathrm{f}$, and $c$ are constants. Approximately, $c$ determines the thickness of the wall and $0 \leq \tau_{0} \leq \tau_\mathrm{f}$ determines the location of the wall. It can be seen that the boundary conditions at the initial and the end points have been guaranteed in the sense that $\phi(0)=\phi_0=0$, $\phi(\tau_\mathrm{f})=1$, and $\dot{\phi}(0)=\dot{\phi}(\tau_\mathrm{f})=0$.

Since $\phi(\tau)$ is a monotone function, we can regard $V(\phi(\tau)) = V(\tau)$ as a function of $\tau$. Then the scalar field equation \eqref{eq:ddphi} becomes an equation for $V$ as follows
\begin{equation}
\dot{V} = \dot{\phi}\ddot{\phi} + 3\frac{\dot{a}}{a} \dot{\phi}^{2}.\label{phibecomeV}
\end{equation}
Therefore, the scalar field $\phi(\tau)$ plays a role of an input and the potential $V(\tau)$ becomes an output in this technique.

\begin{figure}
\begin{center}
\includegraphics[scale=0.7]{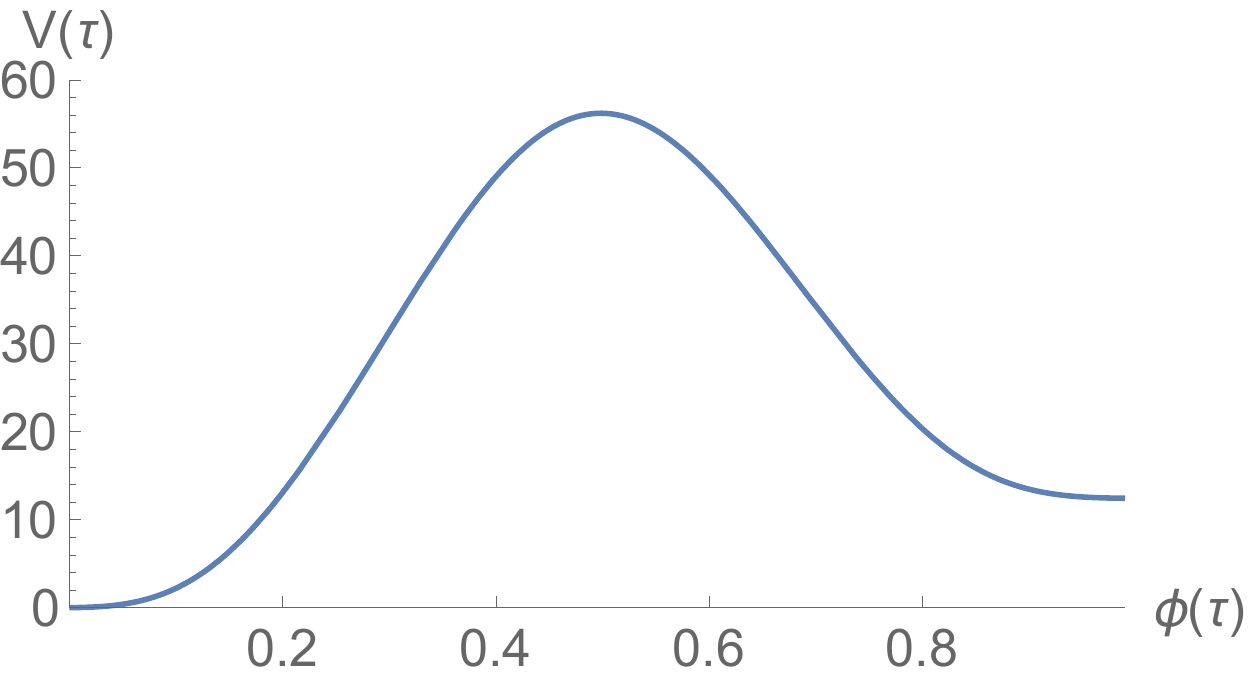}
\includegraphics[scale=0.7]{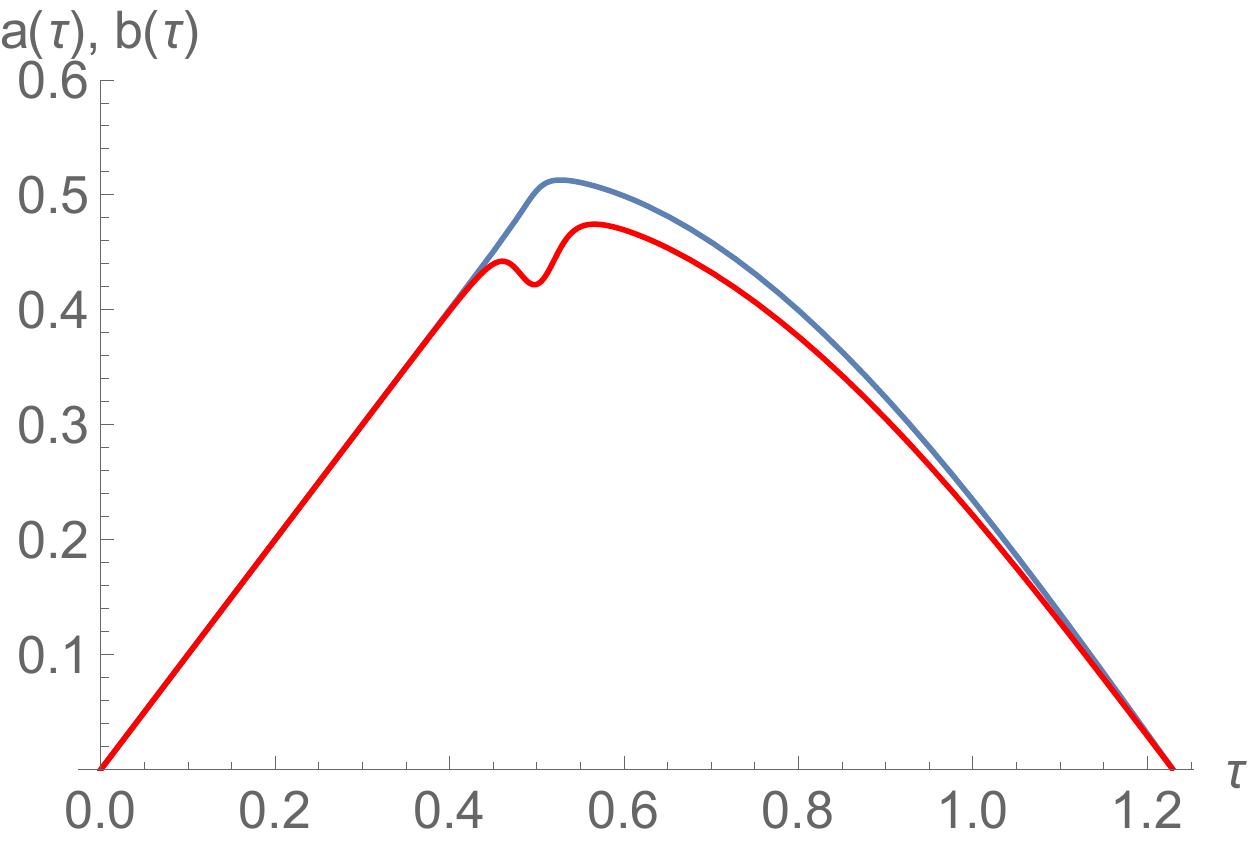}
\includegraphics[scale=0.7]{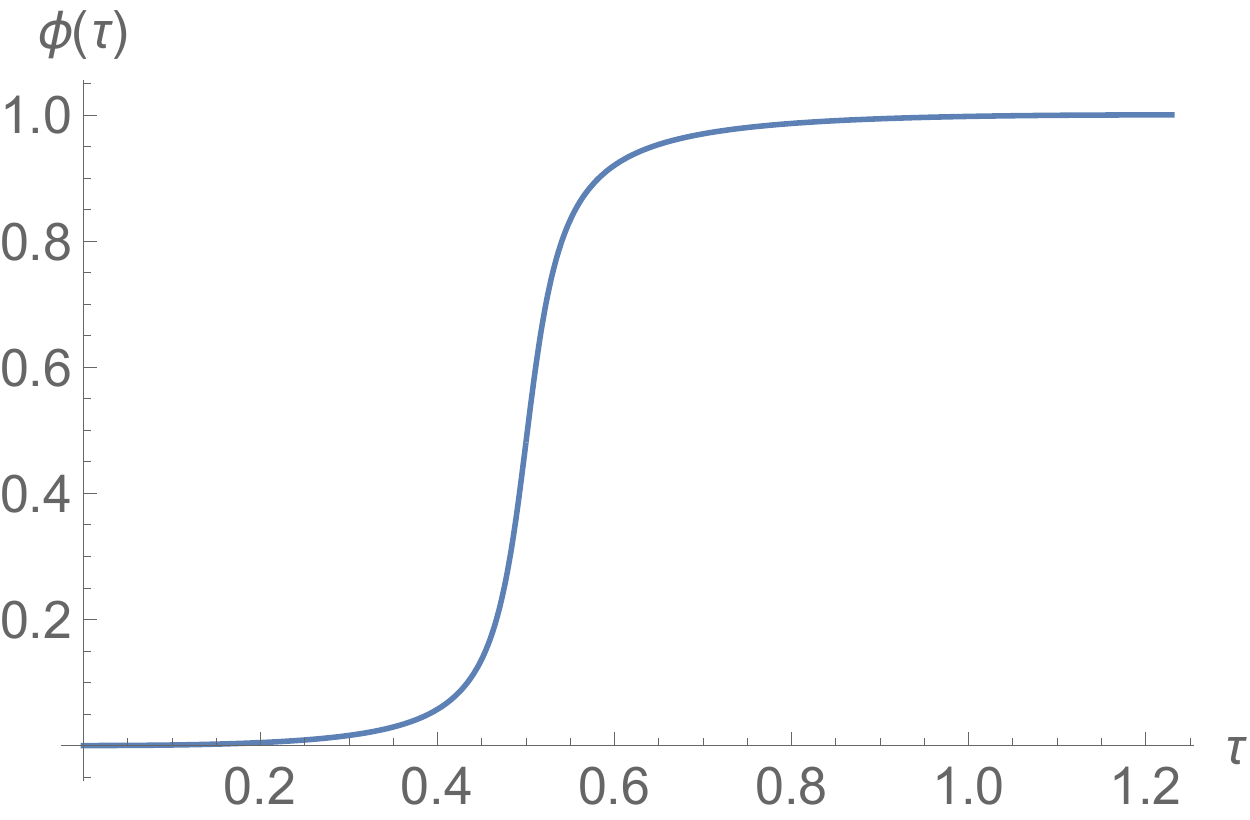}
\caption{\label{fig:pot2}The regular instanton solutions obtained by using the second approach in which a non-trivial scalar field $\phi$ (Eqs.~\eqref{nontriphi1} and \eqref{nontriphi2}) is assumed. Upper: the scalar field potential $V(\phi)$, Middle: the scale factors $a$ (red) and $b$ (blue), Lower: the scalar field $\phi$, for $c = 0.03$, $\tau_{0}=0.5$, $\tau_\mathrm{f} = 1.22953$, $V_{0} = 0$, $\lambda=1$, and $\kappa=0.01$.}
\end{center}
\end{figure}

After replacing the variables $a$, $M$, $\dot{a}$, and $\dot{M}$ with $b$, $V$ and their derivatives, one can solve $b$ as well as $V$ by using Eqs.~\eqref{eqddb} and \eqref{phibecomeV}. Note that the initial conditions that we need to consider here are
\begin{equation}
b(0) = 0\,,\qquad
\dot{b}(0) = \sqrt{\lambda + \kappa V(\phi_{0})}\,,\qquad
V(0) = V_{0}.
\end{equation}
The result can be seen in Figure~\ref{fig:pot2}. The most important result is the \textit{neck} in the scale factor $a(\tau)$ appearing at $\tau\sim\tau_0$. As mentioned previously, the scale factors $a$ and $b$ would behave differently when the kinetic energy of the scalar field is large. In this result, the huge kinetic energy leads to a significant difference between $a$ and $b$. Later, we will show that this \textit{neck} feature can be interpreted as a Lorentzian wormhole when the Euclidean space is transformed into the Lorentzian spacetime after Wick rotation.

Before closing this section, we want to discuss how the fine-tuning problem of the boundary condition can be ameliorated in this approach. Although it turns out that one still needs to assume a fine-tuned $\tau_\mathrm{f}$ in the definition of the scalar field, we have checked from our numerical calculations that different choice of $\tau_\mathrm{f}$ does not change the shape and the structure of the solution much in the sense that the solution still represents a well-structured regular instanton with its real end point different from $\tau_\mathrm{f}$. For example, one can start with an arbitrary $\tau_\mathrm{f}=\tau_\mathrm{f0}$ to find the actual end point $\tau_\mathrm{f1}$ where the scale factors vanish. Then, we redo the numerical calculation by inserting the new end point $\tau_\mathrm{f}=\tau_\mathrm{f1}$ to find a new end point $\tau_\mathrm{f2}$. One repeats the procedures explained above till the input $\tau_\mathrm{f}$ is identical with the real end point. In the first method discussed in subsection~\ref{firstmethod}, the structure of the solution would be drastically different even if $\phi_0$ only deviates slightly  from its fine-tuned value. However, in the second approach, the solution maintains its overall structure when choosing a different $\tau_\mathrm{f}$. This makes the determination of the real $\tau_\mathrm{f}$ much easier.  

\section{Tunneling probabilities}\label{secprob}
\begin{figure}
\begin{center}
\includegraphics[scale=0.7]{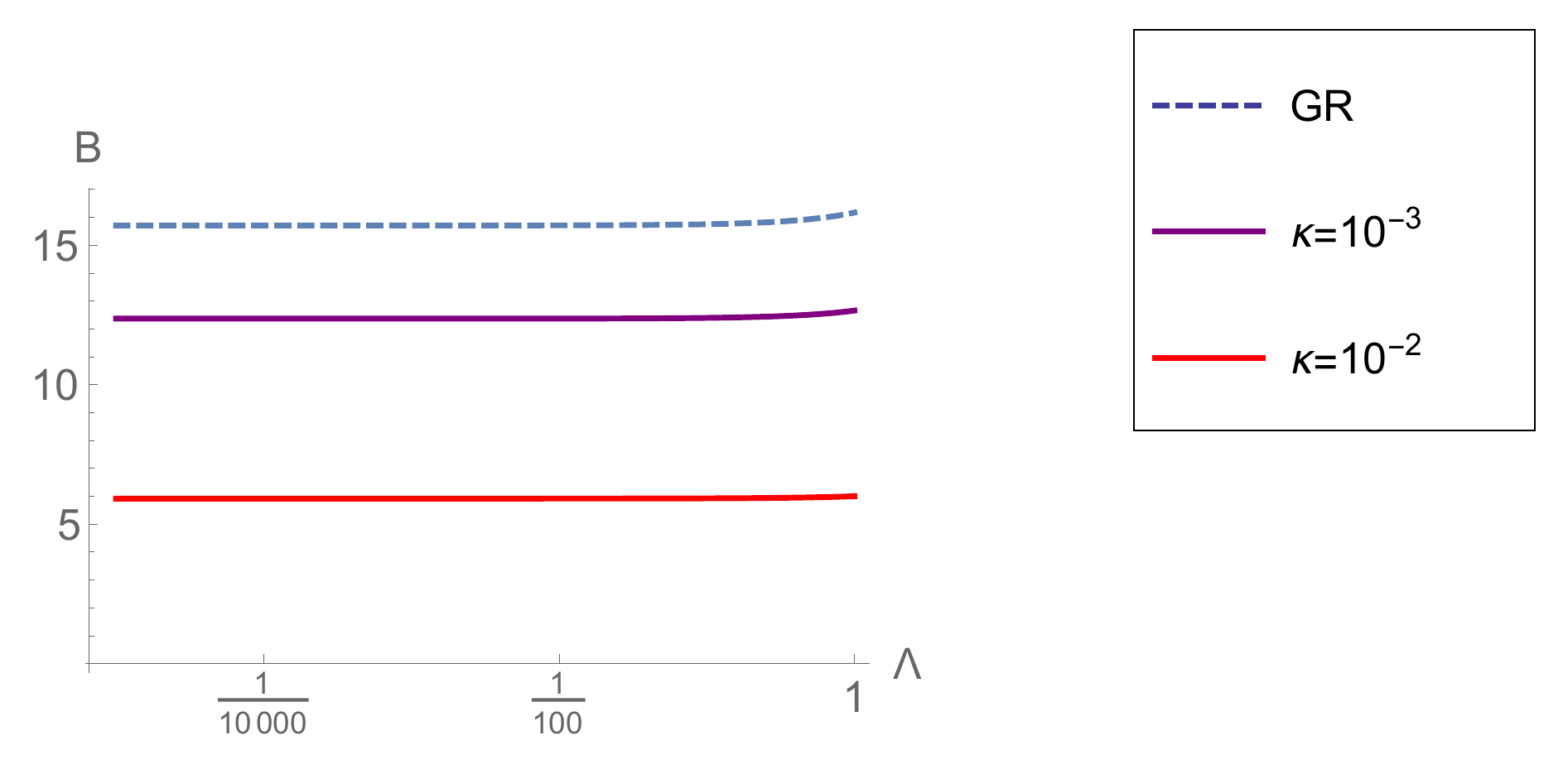}
\includegraphics[scale=0.7]{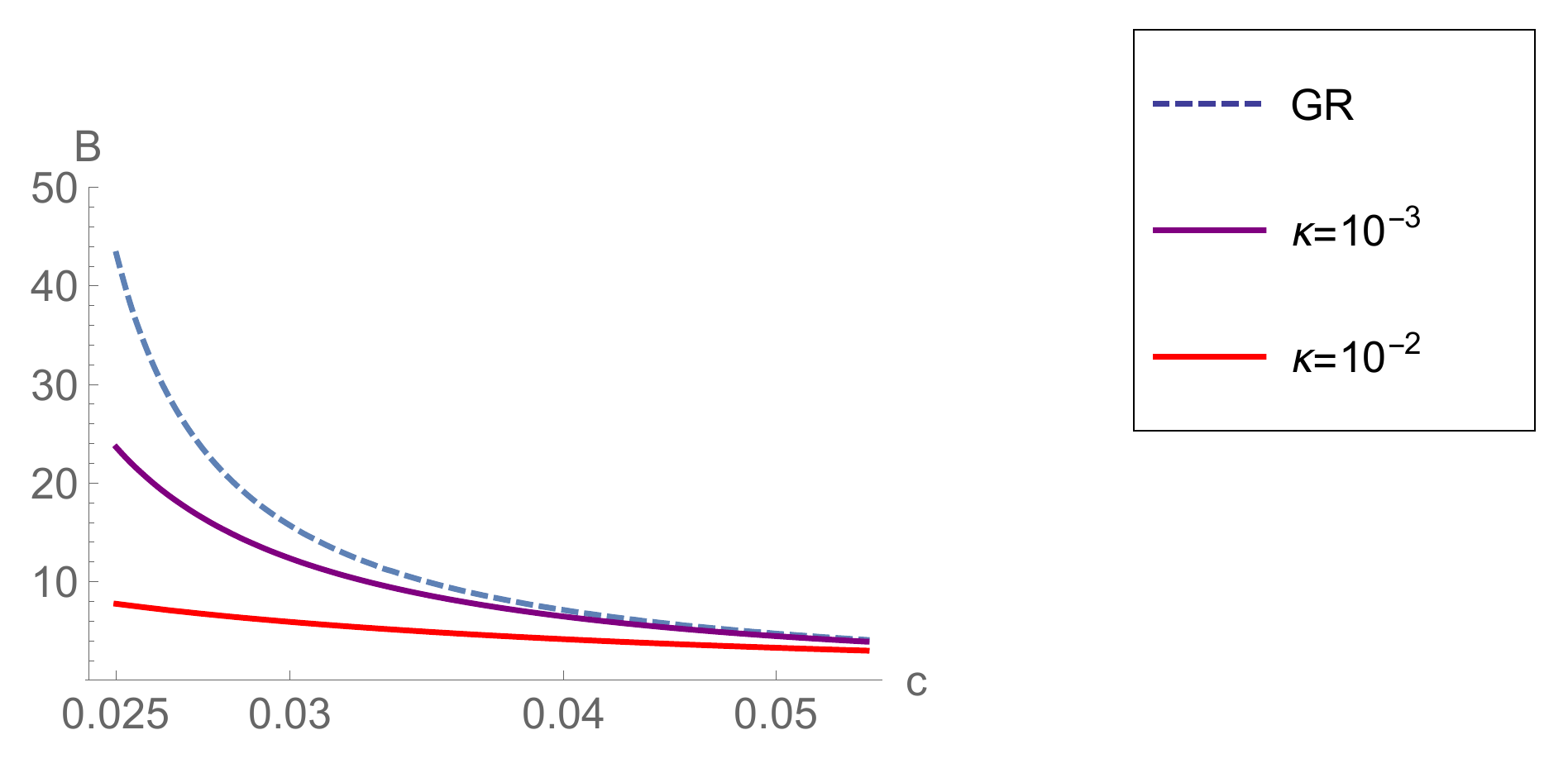}
\caption{\label{fig:probnum}Numerical results of $B$ of the solutions obtained in section~\ref{seciii}. In the upper figure, we fix $c=0.03$, while vary $\kappa$ and the effective cosmological constant $\Lambda$. In the lower figure, we assume a zero $\Lambda$, while vary $\kappa$ and $c$.}
\label{num}
\end{center}
\end{figure}

According to the Euclidean on-shell action given in Eqs~\eqref{2.11} and \eqref{2.12}, the tunneling rate per unit volume is defined by
\begin{equation}
\gamma\propto e^{-B}\,,
\end{equation} 
where the tunneling factor $B$ is
\begin{align}
B&=S_{\mathrm{E}}(\textrm{solution})-S_{\mathrm{E}}(\textrm{false vacuum})\nonumber\\
&=\left\{\frac{2\pi^{2}}{\kappa} \int d\tau a^{3} \left( 1+\kappa\Lambda + \kappa V(\phi) + \kappa \frac{\dot{\phi}^{2}}{2} \right) \left( 1 - \frac{b^{2}}{a^{2}} \right)\right\}+\frac{24\pi^2(1+\kappa\Lambda+\kappa V_\mathrm{F})}{\Lambda+V_\mathrm{F}}.\label{numb}
\end{align}
In the first term, the integration interval is from false vacuum to true vacuum. The last term is the contribution from the false vacuum state. The numerical results of $B$, based on the solution obtained in the previous section, are exhibited in Figure~\ref{num}. 

In the upper figure of Figure~\ref{num}, we fix the parameter $c=0.03$, while vary the EiBI constant $\kappa$ and the effective cosmological constant $\Lambda$. In the lower figure, on the other hand, we assume a zero $\Lambda$, while we vary $\kappa$ and $c$. It can be seen from both figures that the tunneling factor $B$ would decrease with the EiBI coupling constant $\kappa$ in the parameter space we are considering. Furthermore, the probability is almost independent of the effective cosmological constant as long as $\Lambda$ is small. Finally, our numerical results reveal that a larger $c$ would decrease the potential barrier between the two vacua, hence increase the tunneling probability ($B$ becomes smaller).  

\section{Analytical continuation: Time-like wormholes in the physical metric}\label{secvwh}
As we have mentioned, in the EiBI gravity the regular instanton solution would behave significantly differently form its GR counterpart when the kinetic energy of $\phi$ is large. The \textit{neck} in the scale factor $a(\tau)$ shown in Figure~\ref{fig:pot2} reflects this interesting behavior. We will show in this section that this \textit{neck} feature can be interpreted as a wormhole when one considers the materialization of bubble geometries.

\subsection{A Lorentzian wormhole accompanied with the bubble materialization}
The materialization of bubbles can be understood through an analytical continuation of the Euclidean metric
\begin{align}
ds_{g}^{2} &= d\tau^{2} + a^{2}(\tau) d\Omega_{3}^{2}\nonumber\\
&=d\tau^{2} + a^{2}(\tau)\left(d\chi^2+\sin^2\chi d\Omega_2^2\right),\label{metric5.1}
\end{align}
to the Lorentzian spacetime. If we consider the following analytical continuation: $\chi = \pi/2 + iT$, the physical metric \eqref{metric5.1} becomes \cite{Hawking:1998bn}
\begin{eqnarray}
ds_{g,\mathrm{I}}^{2} = d\tau^{2} + a^{2}(\tau) \left( - dT^{2} + \cosh^{2} Td\Omega_{2}^{2} \right).\label{lor1}
\end{eqnarray}
In this coordinate, $0 \leq T < \infty$ is the time-like parameter (constant $T$ surfaces are space-like) and $\tau$ is the space-like parameter (constant $\tau$ surfaces are time-like). Note that $\tau = 0$ and $\tau = \tau_\mathrm{f}$ are null surfaces since $a(0)=a(\tau_\mathrm{f})=0$.

Beyond the null surfaces, one can further analytically continue the spacetime \eqref{lor1} by choosing $\tau = it$, $T = i\pi/2 +\zeta$, and $\alpha(t) = -ia(it)$:
\begin{eqnarray}
ds_{g,\mathrm{II}}^{2} = - dt^{2} + \alpha^{2}(t) \left( d\zeta^{2} + \sinh^{2} \zeta d\Omega_{2}^{2} \right).
\end{eqnarray}
One can do the same thing for $\tau = \tau_\mathrm{f} + it$. This corresponds to an open slicing of the de Sitter space.

In the bubble materialization, the bubble wall is located at $ds^{2}_{g,\mathrm{I}}$. On the wall, there appears a \textit{neck} region where $\dot{a} = 0$ and $\ddot{a} > 0$ according to the solution described in Figure~\ref{fig:pot2}. This means that around the wall, there is a throat in the sense that the physical radius changes from a decreasing phase to an increasing phase. Moreover, the shell dynamics is time-like, hence the location of the throat is time-like. The wormhole structure can be more easily comprehended by considering a constant $T$ surface in $ds_{g,\mathrm{I}}^{2}$. Without loss of generality, we choose the surface $T=0$ and get
\begin{align}
ds_{g,\mathrm{I}}^{2}\big|_{T=0}&=d\tau^2+a^2(\tau)d\Omega_{2}^{2}\nonumber\\
&=\Big(\frac{d\tau}{da}\Big)^2da^2+a^2d\Omega_{2}^{2}.\label{constantwh}
\end{align}
We then compare the above spatial metric with the spatial section of the Morris-Throne wormhole:
\begin{equation}
ds^2_{\mathrm{MT}}=\frac{da^2}{1-B(a)/a}+a^2d\Omega^2_2,
\end{equation}
where $B(a)$ is the shape function of the wormhole. It is well-known that the following two conditions should be satisfied at the throat:
\begin{itemize}
\item[1.] The throat appears at $a=B(a)$.
\item[2.] The flare-out condition: $B(a)-a\frac{dB}{da}>0$.
\end{itemize}
It can be easily seen that the first condition implies $\dot{a}=0$ in the metric \eqref{constantwh}. On the other hand, the flare-out condition at the throat requires
\begin{equation}
\ddot{a}>0.
\end{equation}
Therefore, we can conclude that the \textit{neck} feature can be interpreted as a time-like wormhole (around the shell) in the region of the metric $ds_{g,\mathrm{I}}^{2}$.

\begin{figure}
\begin{center}
\includegraphics[scale=0.7]{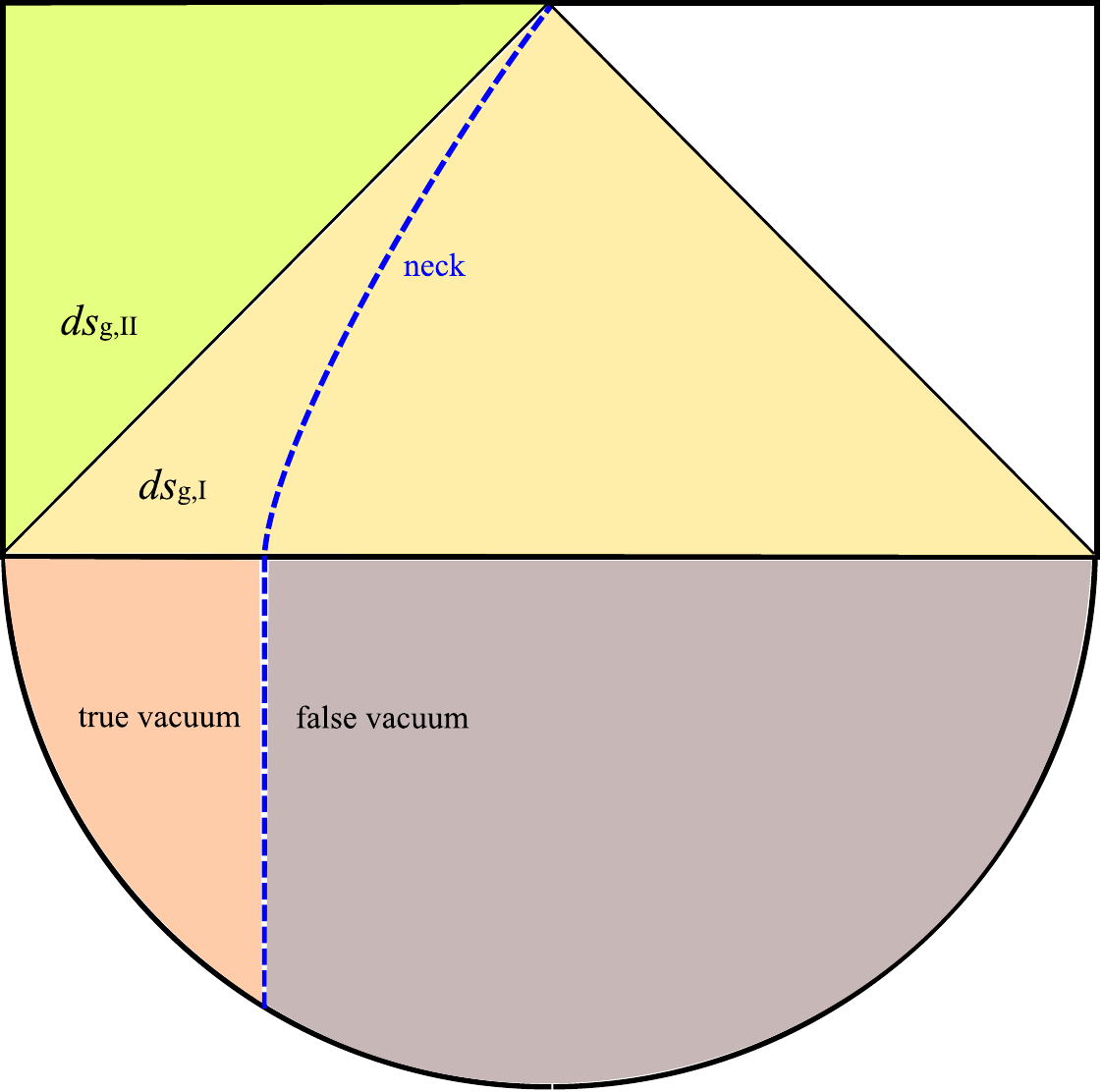}
\caption{\label{fig:picwh}In this figure we show the analytical continuation of the physical metric of the regular instanton in the EiBI theory. The lower half part is the Euclidean section and the upper parts, which are denoted by $ds_{g,\mathrm{I}}$ and $ds_{g,\mathrm{II}}$, are the Lorentzian sections. We analytically continue from the $\chi=\pi/2$ hypersurface of the Euclidean manifold to the Lorentzian manifold $ds_{g,\mathrm{I}}$. The wormhole is at the Lorentzian spacetime represented by $ds_{g,\mathrm{I}}$ and is depicted by the dashed curve.}
\end{center}
\end{figure}

We can apply the same analytical continuation to $ds_{q}^{2}$, but in this case, there is no wormhole. However, local observations will be performed within the physical metric and hence the physical observers will measure a time-like wormhole. The analytical continuation from the Euclidean metric to the Lorentzian spacetime and the wormhole are exhibited in Figure~\ref{fig:picwh}.

\subsection{Null energy condition and null convergence condition}
It is well known that the existence of a Lorentzian wormhole in GR is accompanied by the violation of the null energy condition (NEC). This is because the throat geometry requires some sorts of repulsive features in the spacetime and the violation of the null convergence condition (NCC). In fact, the NCC is satisfied (violated) if and only if the NEC is satisfied (violated) due to the Einstein equation:
\begin{equation}
T_{\mu\nu}n^{\mu}n^{\nu}\ge 0\quad(\textrm{NEC})\qquad\longleftrightarrow\qquad R_{\mu\nu}(g)n^{\mu}n^{\nu}\ge 0\quad(\textrm{NCC}),
\end{equation}
where $n^{\mu}$ is the four velocity of light ray satisfying $g_{\mu\nu}n^{\mu}n^{\nu}=0$. This is exactly the reason why the existence of Lorentzian wormholes in GR requires some exotic matters which violate NEC to support such wormhole geometries. However, in modified theories of gravity such as the EiBI theory that we are considering, it is possible that the NCC is violated while the NEC is satisfied. Therefore, one can construct Lorentzian wormhole solutions without including any exotic matter \cite{Olmo:2013gqa,Olmo:2015bya,Olmo:2015dba,Harko:2013aya,Shaikh:2015oha,Tamang:2015tmd}. Here we will briefly illustrate that the Lorentzian wormhole formed during the bubble materialization is also related to the violation of NCC, even though the NEC is fulfilled.  

\begin{figure}
\begin{center}
\includegraphics[scale=0.7]{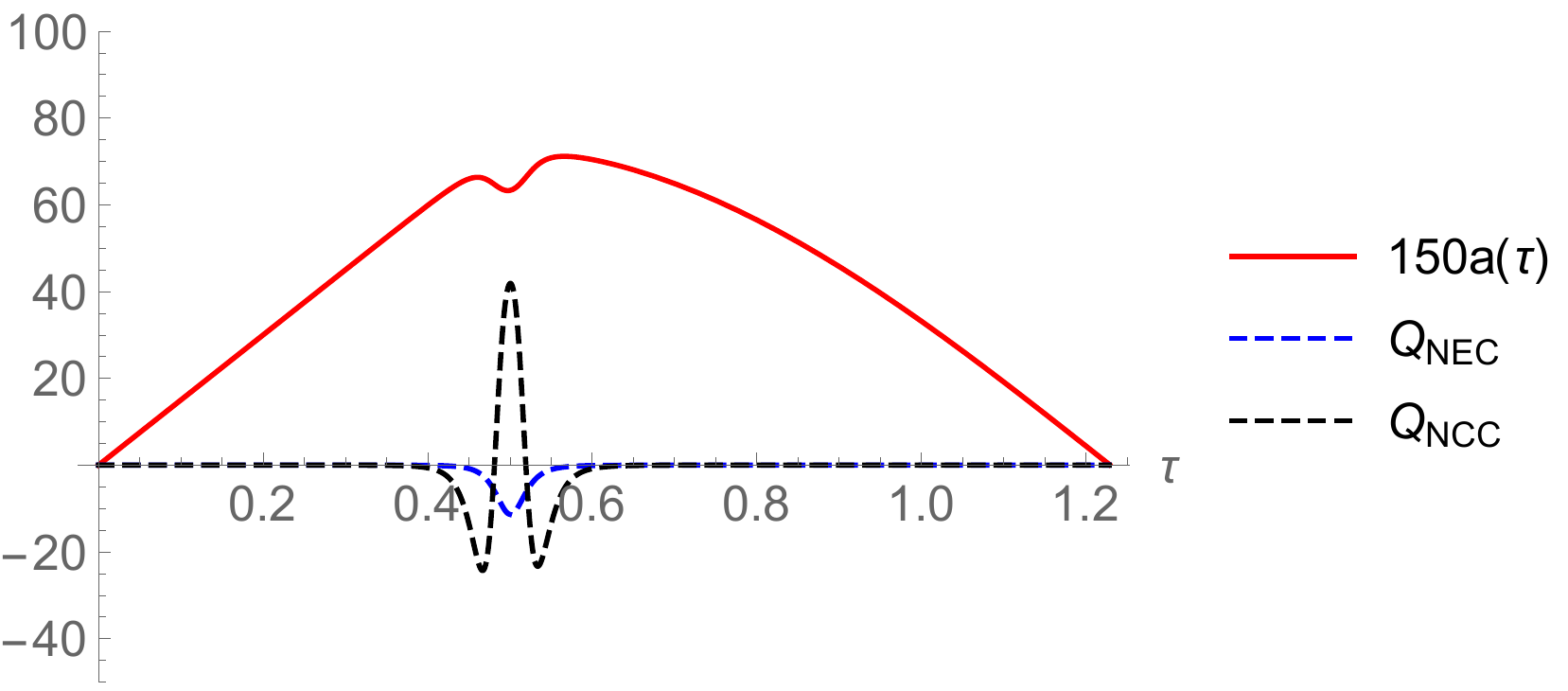}
\caption{\label{fig:probBCC}The behavior of $Q_{\mathrm{NCC}}$ (dashed black), $Q_{\mathrm{NEC}}$ (dashed blue), and the numerical solution of the scale factor $a$ found in Figure~\ref{fig:pot2}. It can be seen that near the wormhole neck, the NCC is violated while NEC is still satisfied.}
\end{center}
\end{figure}

In the Euclidean space, we can define the following two quantities to quantify NCC and NEC \cite{Battarra:2014naa}:
\begin{align}
Q_{\mathrm{NCC}}&\equiv a^2\frac{dH}{d\tau}+1,\\
Q_{\mathrm{NEC}}&\equiv -\frac{a^2}{2}\dot{\phi}^2,
\end{align}
respectively, where $H\equiv\dot{a}/a$. More explicitly, the violation of NCC (NEC) corresponds to a positive $Q_{\mathrm{NCC}}>0$ $(Q_{\mathrm{NEC}}>0)$. Note that $Q_{\mathrm{NCC}}=Q_{\mathrm{NEC}}$ in GR because of the Einstein equation. In Figure~\ref{fig:probBCC}, we exhibit how the physical NCC is violated, even though the NEC is satisfied, near the neck which has been found in Figure~\ref{fig:pot2}.

Note as well that the existence of Enclidean wormholes in GR does not necessarily imply the violation of NEC. However, the wormhole solution that we have found in this paper is a Lorentzian wormhole in the sense that the wormhole geometry appears after Wick rotating the Euclidean space to the Lorentzian spacetime $ds^{2}_{g,\mathrm{I}}$. Therefore, the fulfillment of NEC at the neck does hint toward the novel characteristics of the EiBI theory.

\section{Conclusion}\label{conclu}
In this paper, regular instanton solutions are investigated in the EiBI theory by considering the Euclidean version of the theory with a scalar field $\phi$ minimally coupled to gravity. As opposed to the analysis in Ref.~\cite{Arroja:2016ffm}, in which the singular instanton solutions in the EiBI gravity were discussed, in this paper we study instanton solutions with regular boundary conditions on the two end points. The instanton describes the quantum gravitational tunneling between two vacua and it can be interpreted as the bubble materializing process. When the EiBI coupling constant $\kappa$ vanishes, the theory reduces to GR in which the standard CDL instanton solution is recovered.

We have used two different approaches to obtain the instanton solutions. The first method consists of the preliminary assumption of a double-well potential. This method has been used to confirm the existence of regular instanton solutions in this gravity theory, while it requires a fine-tuned initial value of the scalar field, i.e., $\phi_0$. In the second approach we assume a non-trivial scalar field $\phi(\tau)$ satisfying the boundary conditions at the two end points. We showed that technically the fine-tuning problem  in the first method can be avoided. Furthermore, we calculated the tunneling probability numerically and found that it increases with $\kappa$ in the parameter space of our interest. 

Perhaps the most provocative result in this work is the neck feature of the physical metric $g_{\mu\nu}$ appearing in the Euclidean sector. At the vicinity of the neck, the kinetic energy of the scalar field is large and the effects from the EiBI corrections become significant. It can be seen that the two scale factors $a$ and $b$ behave very differently near the neck (the neck feature appears only in $a$, not in $b$). We showed that when the kinetic energy of the scalar field becomes small (near the vacua), the two scale factors overlap and the theory reduces to GR.

Most importantly, after analytically continuing the Euclidean space to the Lorentzian spacetime, the neck feature can be interpreted as a Lorentzian time-like wormhole around the bubble wall. In addition, this wormhole solution appears in the EiBI theory with a positive Born-Infeld coupling constant $\kappa$ and it does not require the violation of the null energy condition. Due to the fact that the field equation in the EiBI theory is rather different from the Einstein equation, the violation of the convergence condition does not necessarily demand the violation of the energy condition. Furthermore, we find it attractive that a Lorentzian wormhole can be formed via quantum tunneling effects in the EiBI gravity. Similar wormhole creations can be realized in other modified gravity models \cite{Battarra:2014naa,Tumurtushaa:2018agq}. The properties of these Lorentzian wormhole solutions deserve more scrutinies.

\acknowledgments
MBL is supported by the Basque Foundation of Science Ikerbasque. She also would like to acknowledge the partial support from the Basque government Grant No. IT956-16 (Spain) and from the project FIS2017-85076-P (MINECO/AEI/FEDER, UE). CYC and PC are supported by Taiwan National Science Council under Project No. NSC 97-2112-M-002-026-MY3, Leung Center for Cosmology and Particle Astrophysics (LeCosPA) of National Taiwan University, and Taiwan National Center for Theoretical
Sciences (NCTS). PC is in addition supported by US Department of Energy under Contract No. DE-AC03-76SF00515. DY is supported by the Korean Ministry of Education, Science and Technology, Gyeongsangbuk-do and Pohang City for Independent Junior Research Groups at the Asia Pacific Center for Theoretical Physics and the National Research Foundation of Korea (Grant No.: 2018R1D1A1B07049126).

\end{document}